\documentclass[%
 aip,
 amsmath,amssymb,
 reprint,%
author-year,%
]{revtex4-1}

\usepackage{graphicx}
\usepackage{dcolumn}
\usepackage{bm}

\usepackage[utf8]{inputenc}
\usepackage[T1]{fontenc}
\usepackage{mathptmx}
\usepackage{etoolbox}

\usepackage{color}
\usepackage{soul}
\usepackage{xfrac}

\makeatletter
\def\@email#1#2{%
 \endgroup
 \patchcmd{\titleblock@produce}
  {\frontmatter@RRAPformat}
  {\frontmatter@RRAPformat{\produce@RRAP{*#1\href{mailto:#2}{#2}}}\frontmatter@RRAPformat}
  {}{}
}%
\makeatother
\begin{document}
\renewcommand{\arraystretch}{2}

\preprint{AIP/123-QED}

\title[A similarity scaling model for the axisymmetric turbulent jet based on first principles]{A similarity scaling model for the axisymmetric turbulent jet based on first principles}
\author{Preben Buchhave}
    \affiliation{Intarsia Optics, S{\o}nderskovvej 3, 3460 Birkerød, Denmark.}

\author{Clara M. Velte}%
    \email{cmve@dtu.dk}
\affiliation{ Department of Civil and Mechanical Engineering, Technical University of Denmark, Nils Koppels Allé, Building 403, 2800 Kongens Lyngby, Denmark.
}%

\date{\today}

\begin{abstract}
    Similarity scaling, when it can be justified, is a powerful tool for predicting properties of fluid flows and reducing the computational load when using mathematical models. Numerous publications describe different applications of this method, using often different scaling laws with one or more scaling parameters. The justification for these laws is often based on some assumptions or references to experimental results. In this paper, \textit{we base the scaling law on basic physical principles of classical Newtonian physics} (Galilei group) and derive some predictions that we apply to a simple model for the axisymmetric turbulent jet. In a companion paper, we compare these predictions to careful measurements on a free jet in the laboratory and evaluate how far our model predictions are borne out by the experimental results. We have succeeded in obtaining such high measurement quality that we can compute both second and third order statistical functions even far downstream and far-off axis. We can already here reveal that we find very good agreement between a simple one-parameter geometric scaling law derived from the model and numerous first order and higher order statistical results computed from the experimental data.
\end{abstract}

\maketitle

\section{Introduction}
This work presents an alternative derivation of self-similarity properties of a simple model of an axisymmetric jet. Our derivation is based solely on basic symmetry relations inherent in classical Newtonian physics, in contrast to the conventional derivation of the self-similarity relations which is based either on ad hoc assumptions or inspired from results of experiments. Our derivation is of fundamental interest, because it shows that self-similarity for a free flow like the axisymmetric air jet issuing into unrestrained, quiescent air is an inevitable consequence of basic symmetry relations governing the axioms of classical properties of space and time.

Fluid flows that are allowed to develop freely may evolve in a particular manner, called self-preservation or self-similarity, apparently only characterized by the way they are created and sustained, whether it be a boundary layer with one fixed surface, a free jet or a free wake, have been described in numerous publications, see e.g.~\cite{Prandtl1925,von1930mechanische,taylor1939some,zel1937limiting,corrsin1943investigation,batchelor1948energy} for some early descriptions. The theoretical foundation for the concept of self-similarity has largely been based on assumptions, of course motivated by experimental evidence, and references to the commonly accepted theory for self-similarity in general. The theory relating to the free axisymmetric jet, to which we shall limit our description in the following, may be found in textbooks, e.g.~\cite{monin2007statistical,pope2000turbulent,tennekes1972first,frisch1995turbulence} and many others. In these presentations, the presumed similarity behavior regarding various statistical functions is formulated mathematically, and these relations are then inserted into the equations of motion and shown to satisfy these, see for instance~\cite{zel1937limiting,corrsin1943investigation,george1989self}. The actual equation of motion used can either be the full Navier-Stokes equation or some reduced form describing a specific flow, a mean value equation or an equation describing only the fluctuating components. In order to validate these results, it has usually been deemed necessary to compare the derived similarity properties with experimental results.

We think self-preserving flows must be guided by some deep fundamental properties of space and time, so-called symmetries. In the case of most common fluid flows, it is generally agreed that these phenomena can be described in space and time of classical physics, neglecting quantum and relativistic effects and considered a continuum physics phenomenon. This frame of reference is also known as Newtonian physic. For such flows, the importance of self-similarity is, as mentioned, that it can help simplify the mathematical description and reduce the computational load in the application of mathematical engineering models. However, self-similarity is also interesting from a theoretical point of view: What is the physical reason for the development of self-similarity? When will it occur? And to what order of the statistical description of turbulence will it apply?

In the current work, we conclude that first order statistical functions scale according to a single, geometrical scaling factor, and that the same scaling relations with the same scaling factor hold for second and third order statistical functions. This is interesting in relation to the still ongoing discussions in the literature on this issue, where some uncertainty exists as to whether higher order moments scale similarly to first order moments, and whether the scaling of the different order statistical moments relates to the same virtual origin, see e.g.~\cite{carazzo2006route}.

Our analysis does not lead to relations for the first order statistical moments that are different from the ones previously found in standard textbooks, like~\cite{monin2007statistical} or~\cite{pope2000turbulent}. However, our analysis, based on the simple model we use here, concludes that higher order statistical functions have the same scaling behavior as first order functions -- a result which would be of great interest for applications and model building. Not least since the dissipation rate, which is a central parameter in turbulence modeling, can be predicted. Furthermore, our analysis brings the concept of similarity onto a more basic footing and understanding than the conventional theory, which is based on assumptions that must be confirmed by experiments.

We proceed by listing the fundamental classical symmetries of space and time and the corresponding mathematical-physical conservation relations. We then apply these conservation relations to a simple model for a free, high Reynolds number, incompressible turbulent jet with the only constraint being that it is axisymmetric about a particular direction, the jet axis, but without relying on any underlying assumptions of similarity. We derive the expected self-similarity properties of statistical mean values and higher order moments as well as other high order statistical functions and show how these statistical functions are expected to scale downstream and across the jet. We present these expected scaling relations in Table~\ref{tab:1} in section~\ref{sec:PredScalings}. 

In a companion paper,~\cite{Zhuetal}, we describe measurements with a sophisticated laser Doppler anemometer (LDA), see~\cite{Yaacobetal}, in a high Reynolds number jet in air. We present measured results of ensemble averaged spatial (mapped from temporal measurements using the convection record, see~\cite{buchhave2017measurement}) mean values of statistical functions corresponding to the ones described in the present paper. In~\cite{Zhuetal}, we compare the measured results and scaling behavior of the real jet to the ones expected from the theoretical treatment of the simple jet model and discuss the implications of the results of this study to the existing conventions regarding the self-similarity properties of the free, axisymmetric jet.

\section{\label{sec:TheModel}The Model}
We propose a model for a stationary axisymmetric turbulent jet in air, which is as simple as possible: We assume the jet expands freely without disturbing effects from the environment. We also neglect the effects of gravity. We choose to disregard the fact that the interface between the turbulent jet and the surrounding air is highly convoluted and forms a fractal surface. Instead, we consider a smooth interface to the surrounding air. We assume that we can make the step to a more realistic jet later with reference to the known fractal dimension of the jet surface, see e.g.~\cite{sreenivasan1986fractal} and~\cite{da2011intense}.

\begin{figure}[b]
    \centering
    \includegraphics[scale=0.98]{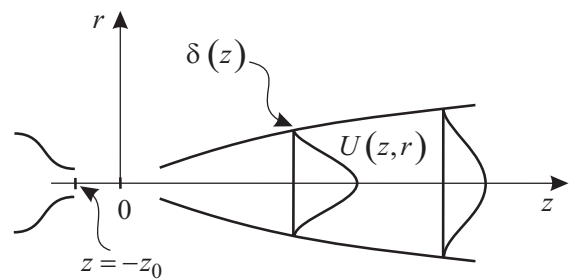}
    \caption{Simple jet model with arbitrary spreading rate.}
    \label{fig:1}
\end{figure}

We limit our considerations to a non-swirling jet with constant density, $\rho$, and consider only the part of the jet which is completely free of influence from the surroundings. Thus, we exclude the initial developing part of the jet, which is directly influenced by the jet exit conditions. Initially, we do not make any assumptions regarding the form of the model jet, whether it be straight conical or changes shape downstream, see Figure~\ref{fig:1}. We also assume that the mean velocity profiles of the jet are integrable, since the radial profiles must, to be physical, at any downstream position, govern the finite flow of mass, momentum and energy. We may then take the width of the jet, $\delta (z)$, to be the radius to a point which encompasses a certain fraction, for example 50\%, of the mean momentum rate.

We apply a cylindrical coordinate system with streamwise coordinate, $z$, radial component, $r$, and azimuthal angle, $\varphi$. The origin of the coordinate system is assumed to be the point of issue of a possible self-similarity to be discussed in the following. The distance from the origin to the jet exit, $z_0$, is determined by the initial conditions and the developing part of the jet, which is not covered by this simple jet model and must be determined by experiment.

In the following, we assume that we measure the axial instantaneous velocity component, $u(z,r,\varphi,t)$, in a time record, $T_s$, sufficiently long to provide accurate statistical results. However, throughout the current work, we assume that we have mapped the temporal records $u(z,r,\varphi,t)$ to spatial records $u(z,r,\varphi,s)$, where $s$ is the fluid path length along the instantaneous convection direction through the measurement point, the so-called ``convection record'' described in~\cite{buchhave2017measurement}. By this method, we avoid the ``sweeping effect'' and problems related to temporal sampling bias. We then describe the mean velocity profile of the axial velocity component as the average over a record, $U(z,r)=\overline{u(z,r,\varphi,s)}$, where average over the measured record is indicated by an overbar. Due to the axisymmetry of the underlying geometry, the dependency on the tangential direction, $\varphi$, vanishes for the average profiles. 

To derive the properties of this simple model, we apply some basic symmetry properties of space and time in which this model jet exists, the Galilei group, see e.g.~\cite{frisch1995turbulence}, namely,
\begin{itemize}
    \item Space translation symmetry
    \item Time translation symmetry
    \item Rotational symmetry
\end{itemize}
In addition, we assume a Newtonian fluid described fully by
\begin{itemize}
    \item Navier-Stokes Equation (NSE)
\end{itemize}
As is well known from Noether's theorem,~\cite{Noether1918}, these symmetries lead to corresponding basic mathematical physics conservation laws:
\begin{itemize}
    \item Conservation of momentum
    \item Conservation of energy
    \item Conservation of angular momentum
\end{itemize}
In addition, we know that if a constant Reynolds number along the jet axis can be shown to exist, we can employ (see e.g.~\cite{monin2007statistical}):
\begin{itemize}
    \item Reynolds number similarity,
\end{itemize}
which, as we shall see, allows us to derive several similarity properties for the free jet.

We need an objective definition for the large velocity scale, $\mathcal{U}(z)$, and a corresponding definition of the large spatial scale, $\mathcal{L}(z)$, with the aim of investigating whether the large-scale Reynolds number, $Re_{\mathcal{L}}(z) = \mathcal{L}(z) \mathcal{U}(z) / \nu$, is constant along the axis of the jet, with $\nu$ denoting the kinematic viscosity of the working fluid.

The velocity scale, $\mathcal{U}(z)$, must uniquely characterize the jet and have the dimension of velocity. It must be independent of the shape of the unknown velocity profile, but could conceivably be related to the integrated velocity or momentum flux across the local cross section. This definition will then be independent of the shape of the particular local velocity profile because of momentum conservation and will thus not risk to implicitly impose self-similarity. 

Assuming the jet to be fed by a source of constant rate of momentum in the $z$-direction, for example a top-hat profile at the exit of the jet, we conclude from momentum conservation and~\cite{Noether1918}, that the instantaneous momentum flux integrated over any cross section of the jet must be identical to the initial rate. This will of course then also hold true for the average momentum flux. We can then write for the constant momentum rate along the $z$-axis, $\dot{B}$, 

\begin{equation}
    \dot{B} = \rho \int_{0}^{\infty}2 \pi r\, U^2(z,r) \, dr = \mathrm{constant}
\end{equation}
We can define the large spatial scale, $\mathcal{L}(z)$, as the radial distance enclosing a certain fraction of the momentum rate, for example half the momentum rate: 
\begin{equation}
    \frac{1}{2}\dot{B} \equiv \rho \int_{0}^{\mathcal{L}(z)}2 \pi r\, U^2(z,r)\, dr.
\end{equation}
We then define the large velocity scale, $\mathcal{U}(z)$, so that the (constant) momentum transfer rate expressed by the large scales equals the total momentum rate:
\begin{equation}
    \rho \pi \mathcal{L}^2(z) \mathcal{U}^2(z) \equiv \dot{B} = \rho \int_{0}^{\infty}2 \pi r\, U^2(z,r)\, dr.
\end{equation}
We then obtain the expression for the large velocity scale:
\begin{equation}
    \mathcal{U}(z) = \sqrt{\frac{1}{\pi \mathcal{L}^2(z) } \int_{0}^{\infty}2 \pi r\, U^2(z,r) \, dr}
\end{equation}

This definition is based on a conserved quantity, $\dot{B}$, independent of the shape of the actual velocity profile. Then the product $\mathcal{L}(z)\mathcal{U}(z)$, and, consequently, the large-scale Reynolds number $Re_{\mathcal{L}}(z) = \mathcal{L}(z)\mathcal{U}(z)/\nu$, must be constant along the axis of the jet. This result is thus based on a conserved quantity, the rate of momentum, and is independent on the actual shape of the jet.

Thus, at different positions along the axis in the fully developed jet region, say at $z_1$ and $z_2$, the non-dimensional Navier-Stokes equation with the non-dimensional variables, $z_1^{\ast} = \frac{1}{\mathcal{L}(z_1)}z_1$ and $z_2^{\ast} = \frac{1}{\mathcal{L}(z_2)}z_2$, will have identical solutions and hence $z_1^{\ast} = z_2^{\ast} \equiv z^{\ast}$. Consequently, the non-dimensionalized velocity profile, and with that the jet width (regardless of definition applied), will be identical along the jet axis. We can then conclude after applying the reverse transformation equations, $z = \mathcal{L}(z) z^{\ast}$ and $r = \mathcal{L}(z) r^{\ast}$, that the jet width will be proportional to $z$: $\delta (z) \propto \frac{1}{z^{\ast}}z$, and the jet will expand linearly from a common center of origin.

We also obtain $\mathcal{U}(z) = \frac{\nu Re_{\mathcal{L}}(z)}{\mathcal{L}(z)} = \nu Re_{\mathcal{L}}(z) \frac{z^{\ast}}{z}$. Thus, the large velocity scale, $\mathcal{U}(z)$, is inversely proportional to $z$. Since solutions to the Navier-Stokes equation, and hence the non-dimensional velocity profiles, are identical at different axial distances, we may introduce a non-dimensional velocity, $U^{\ast}\left ( r^{\ast} \right ) = \frac{1}{\mathcal{U}(z)}U(z,r)$.  Thus, the velocity profile of the radial velocity component is also expected to show the same single parameter similarity scaling as the axial velocity component. 

Apparently, self-similarity is a consequence of basic physical symmetries, at least for this simple jet model. In addition, the fact that we assume rotational symmetry about a line in space, corresponding to the axis of the jet, means that we can also invoke conservation of angular momentum around that axis, which simply means that a non-swirling jet at the exit will remain non-swirling throughout. 

Alternatively, one could instead have defined $\mathcal{L}(z)$ in the classical manner as the radial distance to a point of the transverse mean velocity profile with a certain value, for example 50\%, of the centerline velocity. We could then choose a value of $\mathcal{U}(z)$ to provide us with a constant Reynolds number, $Re_{\mathcal{L}}(z) = \frac{\mathcal{L}(z) \mathcal{U}(z)}{\nu}$. This would allow us to claim Reynolds number similarity and consequently self-similarity for the jet. The difference from our method might seem small, but it is significant. The second method would amount to implicitly assume or define self-similarity, since one would imply a basic underlying shape of the velocity profiles, whereas our method relies on a conserved quantity, the rate of momentum transport. 

Instead, with our suggestion, the large scales are defined such that they provide the total constant momentum transport, $\dot{B} = \rho \pi \mathcal{L}^2 (z) \mathcal{U}^2 (z)$. It then follows that $Re_{\mathcal{L}} (z) $ must be constant. Reynolds number similarity, and thus self-similarity for the jet, follows. This is the main point of the present work.

\section{\label{sec:FurtherSSprops}Further self-similarity properties for the jet model}
As the jet expands downstream, the mass of air in motion increases downstream by the process of entrainment, causing the mass flow rate to increase, see e.g.~\cite{ricou1961measurements}, who provide excellent experimental evidence. 

We now apply our method to entrainment. This mass influx is supplied by an inward flow from the surrounding still fluid. The mass flow rate integrated over the jet cross section at an axial distance, $z$, is 
\begin{equation}
    \dot{m} = \rho \int_0^{\infty} 2 \pi r \, U(z,r)\, dr. 
\end{equation}
Introducing the non-dimensional variables, we obtain
\begin{equation}
    \dot{m} = \rho \mathcal{L}^2(z) \mathcal{U} (z) \int_0^{\infty} 2 \pi r^{\ast} \, U^{\ast}(r^{\ast})\, dr^{\ast}. 
\end{equation}

As the integral is constant and given the scalings of  $\mathcal{L}(z)$ and $\mathcal{U}(z)$, we see that the mass flow rate increases proportional to the downstream distance, $z$. The entrainment, understood as the mass flow rate per unit downstream distance, is thus constant, independent of the distance from the origin.

We can now use conservation of energy for the fully developed jet in the inertial range to predict some second order statistical properties of the jet. Energy flux, or the total energy transport rate, are not conserved quantities in a real jet; the kinetic energy transport decreases downstream due to an imbalance between production, dissipation etc. Furthermore, the total energy balance between advected kinetic energy, kinetic energy production, dissipation and pressure strain rate contribution depend on the downstream and radial position in the jet. 

We will therefore herein concentrate on the spectral distribution, the so-called power spectral density (PSD), of the turbulent kinetic energy in the \textit{inertial range}. In our case, energy conservation simply means that the total kinetic energy in a slice of the jet of a fixed thickness $\Delta$ does not increase or decrease along any distance $z$ downstream, and that the spectral distribution has reached a stable form characteristic for the fully developed jet. We shall continue with the PSD expressed in Fourier space and investigate, based on the simple model, how to describe the scaling in the ordinate and abscissa direction.

For a given spatial window, $L_s$, the power spectral density is defined as
\begin{equation}
    F(z,r,\varphi,k) = \frac{1}{L_s} | \hat{u}'(z,r,\varphi,k) |^2 
\end{equation}
where the Fourier transform of the velocity fluctuations is given by:
\begin{equation}
    \hat{u}'(z,r,\varphi,k) = \int_{-\infty}^{\infty} e^{-iks} W(s) u'(z,r,\varphi,s)\, ds 
\end{equation}
$W(s)$ is a window function that is unity over the length of the record $L_s$ and zero elsewhere, and the prime indicates the fluctuating part of the velocity. $k$ is the wave number describing the spectrum of the spatial structures convected through the measurement point and is thus also subject to similarity scaling. 

We can now write the kinetic energy $E_{kin}$ of a cross-sectional disc of the jet of fixed thickness $\Delta$ at the axial distance $z$ in terms of the power spectral density, $F(z,r,\varphi,k)$, integrated over the volume elements, $dV = 2 \pi r\, dr \, \Delta$, of the section and over all wave numbers, $k$: 	
\begin{equation}
    E_{kin} = \frac{1}{2}\rho \int_{V} \left (  \int_{k=0 }^{\infty} F(z,r,\varphi,k)\, dk \right ) \, dV. 
\end{equation}
Introducing the scalable, normalized coordinates to the volume element, we obtain
\begin{equation}
    E_{kin} = \frac{1}{2}\rho \mathcal{L}^2(z) \int_{r^{\ast}=0}^{\infty} \left (  \int_{k=0 }^{\infty} F(z,r,\varphi,k)\, dk \right ) \, 2 \pi r^{\ast}\, dr^{\ast}\, \Delta. 
\end{equation}

Since $E_{kin}$ is constant in slices of fixed thickness $\Delta$, both $F$ and $k$ must scale inversely proportional to the distance from the virtual origin: $F \propto z^{-1}$ and $k \propto z^{-1}$. The integral of $F$ over all wave numbers then scales as the mean square velocity (or the velocity variance), inversely proportional to the second power of $z$, as expected.

The scaling of the velocity variance can also be immediately identified from its form (velocity squared): $\overline{u'^2}\propto z^{-2}$. This also results in that the root-mean-square (rms) velocity, $\sqrt{\overline{u'^2}}$, is proportional to $z^{-1}$, and, in the same manner as for the mean velocity, decreases as one over $z$. The turbulence intensity along the jet axis is thus expected to be constant.

Since the position of the jet exit, $-z_0$, is unknown and in fact has to be determined from experiments for each specific jet flow separately, we can conclude that for flow statistics such as the power spectral density, the abscissa $k$ and ordinate $F$ both scale from the jet virtual origin and not the jet exit. 

\section{\label{sec:PredScalings}Predicted scaling of higher order statistical functions}
Knowing the scaling behavior of the jet width, the velocity scale and the size of spatial structures as a function of axial distance, we can now predict the scaling behavior of other statistical quantities based on the dimensions of their respective expressions. This includes dynamic moments, such as structure functions as well as for length scales such as the integral length scale, the Taylor microscale and the Kolmogorov microscale. 

In the following, we envision the $z$-component of velocity measured at the axis and suppress the other coordinate directions. The second order structure function is given by $S_2(l) = \overline{\left ( u(z,r,\varphi,s+l) - u(z,r,\varphi,s) \right )^2 }$, where $l$ is a spatial increment in the axial direction and $s$ is the convection record coordinate. Since the data points in $s$ are spatially separated, we find directly the spatially averaged structure function. This function has the dimension of velocity squared, and we can thus expect it to scale along the ordinate as $z^{-2}$. The third order structure function is given by $S_3(l) = \overline{\left ( u(z,r,\varphi,s+l) - u(z,r,\varphi,s) \right )^3 }$. It has the dimension of velocity to the third power and is therefore expected to scale as $z^{-3}$. The size of the spatial structures $l$, i.e. the abscissa in both cases, is expected to scale proportional to $z$. Thus, the slope of the third order structure function near the origin (defined as $|dS_3/dl|$) scales with the distance $z$ to the minus fourth power. 

The average dissipation rate, $\overline{\varepsilon} = \frac{5}{4} \left | \frac{dS_3 (l)}{dl} \right |$, is related to the slope of $S_3(l)$ by Kolmogorov's Four-Fifth Theorem, and thus is expected to also scale as $z^{-4}$. This can also be deduced from the definition using the scaling of the velocity and length scales, $\overline{\varepsilon} = 2 \nu \overline{s_{ij} s_{ij}} \propto z^{-4}$, where the strain rate tensor $s_{ij}$ consists of spatial velocity derivatives, or even from the common dissipation rate estimator, $\overline{\varepsilon} \propto \overline{\left | u' \right |^3} / \ell \propto z^{-4}$, where $\ell$ is the `pseudo' integral length scale. 

The spatial Kolmogorov microscale is computed from the dissipation by $\eta = \left ( \frac{\nu^3}{\overline{\varepsilon}} \right )^{1/4}$ and has the dimension of length. The spatial Taylor microscale is given by $\lambda = \sqrt{\overline{u'^2} / \overline{\left ( \frac{du'}{dl} \right )^2} }$, and the integral length scale is given by $I = \int_0^{\infty} C(l)\, dl $, where $C(l)$ is the spatial autocorrelation function. All these scales are, accordingly, expected to be proportional to the axial distance from the virtual origin, $z$.

Table~\ref{tab:1} lists the definitions of the statistical quantities, their dimensions, and the expected powers of the scaling parameter, $z$, in the direction of the abscissa and the ordinate when the statistical functions are viewed as graphs in agreement with the experimental plots in the companion paper,~\cite{Zhuetal}. In the table, we have suppressed the coordinates $(z,r,\varphi)$ in the higher order statistical functions.

\begin{table*}
    \caption{Predicted scaling factors of statistical functions.}
    \label{tab:1}
    \centering
    \vspace{0.1cm}
    \begin{tabular}{c|c|c|c|c}
    \hline \hline 
    \bf{Statistical Function} & \bf{Definition} & \bf{Dimension} & \bf{Ordinate Scaling Factor} & \bf{Abscissa Scaling Factor} \\ [0.5ex]
    \hline 
    Axial Distance & $z$ & $L$ & $+1$ &  \\
    Jet Width & $\delta (z)$ & $L$ & $+1$ &  \\
    Top Angle & $\theta_0 = \arctan \left ( \frac{\delta (z)}{z} \right ) $ &  &  &  \\
    Mean Velocity & $U = \overline{dz / dt}$ & $LT^{-1}$ & $-1$ &  \\
    Mean Square Velocity & $\overline{u^2}$ & $L^2T^{-2}$ & $-2$ &  \\
    Velocity Variance & $\overline{u'^2}$ & $L^2T^{-2}$ & $-2$ &  \\
    Turbulence Intensity & $T = \sqrt{\overline{u'^2}}/U$ &  &  &  \\
    Wave Number & $k$ & $L^{-1}$ & $-1$ & \\
    Power Spectral Density & $F(k) = \frac{1}{L_s} \overline{ \left | \hat{u}' (k) \right |^2}$ & $L^3T^{-2}$ & $-1$ & $-1$ \\
    Autocovariance Function & $R(l) = \overline{u'(s)u'(s+l)}$ & $L^2T^{-2}$ & $-2$ & $+1$ \\
    Autocorrelation Function & $C(l) = \overline{u'(s)u'(s+l)} / \overline{u'(s)^2}$ &  &  & $+1$ \\
    Integral Length Scale & $I = \int_0^{\infty} C(l) \, dl$ & $L$ & $+1$ & \\
    Taylor Microscale & $\lambda = \sqrt{\sfrac{\overline{u'^2}}{\overline{\left ( \frac{du'}{dl} \right )^2} }}$ & $L$ & $+1$ & \\
    Kolmogorov Microscale & $\eta = \left ( \frac{\nu^3}{\overline{|u'|^3} / \ell} \right )^{1/4} $ & $L$ & $+1$ &  \\
    Second Order Structure Function & $S_2(l) = \overline{\left ( u(s+l) - u(s) \right )^2 }$ & $L^2T^{-2}$ & $-2$ & $+1$ \\
    Third Order Structure Function & $S_3(l) = \overline{\left ( u(s+l) - u(s) \right )^3 }$ & $L^3T^{-3}$ & $-3$ & $+1$ \\
    $S_3$ Slope & $\alpha  =  \left | \frac{d S_3 (l)}{dl} \right | $ & $L^2T^{-3}$ & $-3$ & $+1$ \\
    Mean Dissipation Rate & $\overline{\varepsilon} = \frac{5}{4} \left | \frac{d S_3 (l)}{dl} \right | $ & $L^2T^{-3}$ & $-3$ & $+1$  
    \end{tabular}
\end{table*}

\section{\label{sec:Discussion}Discussion}
In Table~\ref{tab:1}, we have presented how the statistical quantities scale as a function of the distance from the virtual origin. The intention is to give insights into the physical processes in the jet and how the spatial scales and the mean and fluctuating velocity depend on the axial coordinate $z$. However, since both spatial scale and mean and fluctuating velocities all scale with the axial distance, we may conclude that all the statistical quantities mentioned in Table~\ref{tab:1} scale with a single, purely geometrical scaling factor, namely the distance along the axis of the jet from the virtual origin. This of course based on a simple model of a fully developed jet, dominated by inertial forces and described by the smooth shape of its mean velocity distribution.

The derivation of these results is entirely based on the symmetries of classical physics and the acceptance of Newton's second law (Navier-Stokes' equation) as the axiomatic foundation for the description of classical dynamic processes. It remains, of course, to be seen if these conclusions can be verified by accurate, unbiased measurements in a carefully prepared free, axisymmetric jet. 

We may here again refer to our companion paper,~\cite{Zhuetal}, which describes measurements downstream in a high Reynolds number free jet from 30 to 100 jet orifice diameters. These measurements were performed with a highly accurate laser Doppler anemometer (LDA), capable of measuring in high turbulence intensity including back-flow. The LDA could measure time records at a high sample rate, allowing the computation of the high end of frequency spectra and high-resolution correlation- and structure functions. A special technique based on the measurement of particle transit time (also called residence time) allowed the computation of the instantaneous magnitude of the three-dimensional convection velocity and thereby allowed all statistical functions to be computed as spatial averages, see~\cite{buchhave2017measurement}.

With reference to~\cite{Zhuetal}, we may mention some results and conclusions from that paper that align with the theoretical results obtained herein:
\begin{itemize}
    \item It is concluded that all the first, second and third order statistical functions listed in Table~\ref{tab:1} can be scaled according to a single geometrical scaling factor, namely the distance from a single virtual origin for the fully developed jet.
    \item The virtual origin for all the mentioned statistical functions is determined to be, within experimental accuracy, the same for all the orders, namely approximately five orifice diameters downstream from the jet orifice.
    \item The scaling behavior of static and dynamic moments corresponds to the factors listed in Table~\ref{tab:1}.
\end{itemize}

\section{\label{sec:Conclusion}Conclusions}
In conclusion, it appears that the scaling behavior and self-similarity properties of the free, round jet can be predicted from a simple model without resort to any implicit assumptions of similarity. The derivation of the jet properties is based solely on classical symmetry properties of space and time in a Newtonian physics environment. The scaling behavior of the statistical functions can be predicted with reference to their dimensions. We do not find that any spatial scale, for example the Taylor microscale, plays a special role as suggested by many authors. Instead, the jet development and its statistical quantities are all defined by a single geometrical scaling factor and a single virtual origin.

\section*{Acknowledgements}
Professor Poul Scheel Larsen is acknowledged for helpful discussions.

CMV acknowledges financial support from the European Research council: This project has received funding from the European Research Council (ERC) under the European Unions Horizon 2020 research and innovation program (grant agreement No 803419). 

PB acknowledges financial support from the Poul Due Jensen Foundation: Financial support from the Poul Due Jensen Foundation (Grundfos Foundation) for this research is gratefully acknowledged.

\section*{Declaration of interest}
The authors report no conflicts of interest. 

\section*{Data availability}
Data sharing is not applicable to this article as no new data were created or analyzed in this study.


\nocite{*}
\section*{References}
\bibliography{MainPart1}

\end{document}